# Superconducting Nanocomposites: Enhancement of Bulk Pinning and Improvement of Intergrain Coupling

Ruslan Prozorov, Tanya Prozorov and Alexey Snezhko

*Abstract*—Heterogeneous sonochemical method was applied for synthesis of novel superconducting nanocomposites consisting of magnetic (and/or nonmagnetic) *nanoparticles embedded into the bulk of ceramic superconductors*. In addition to *in-situ* production of the efficient pinning centers, this synthesis method considerably improves the interbrain coupling. Significant enhancement of the magnetic irreversibility is reported for $Fe_2O_3$ nanoparticles embedded into the bulk of $MgB_2$ superconductor. Nonmagnetic $Mo_2O_5$ nanoparticles also increase pinning strength, but less than magnetic $Fe_2O_3$. Detailed magnetization and electron microscopy characterization is presented. Theory of bulk magnetic pinning due to ferromagnetic nanoparticles of finite size embedded into the superconducting matrix is developed.

## I. Introduction

CRITICAL current density is the key parameter determining useful applications of superconductors [1]. Although theoretically it is much easier to deal with perfect crystals, they have little practical use. Thin films are capable of carrying large current *densities*, but the total current remains very small. Therefore, the only option for power applications so far is the bulk form of superconductors. Indeed, it is desirable to use superconductors with elevated transition temperatures. Unfortunately, high-$T_c$ superconductors are granular ceramics and modification of their morphological and chemical properties is much more difficult compared to conventional metallic alloys.

Among variety of methods developed to enhance pinning in

Manuscript received October 4, 2004. This work was supported in part by the donors of the American Chemical Society Petroleum Research Fund and USC Research & Productive Scholarship Award.

R. Prozorov is with the Department of Physics & Astronomy, University of South Carolina, Columbia, SC 29208 USA (corresponding author, phone: 803-777-8197; fax: 803-777-3065; e-mail: prozorov@sc.edu).

T. Prozorov is now with the Department of Chemical Engineering, University of South Carolina, Columbia, SC 29208 USA (e-mail: prozorot@engr.sc.edu).

A. Snezhko was with the Department of Physics & Astronomy, University of South Carolina, Columbia, SC 29208 USA. He is now with the Materials Science Division, Argonne National Laboratory, Argonne, IL 60439 USA (e-mail: snezhko@anl.gov).

superconductors, only few can be used for bulk ceramic materials. Conventional mechanical cold-working is not applicable due to brittleness; irradiation with heavy ions is inefficient due to small ion penetration depth; thermal cycling does not generally lead to the same results as in alloys (e.g. Nb-Ti). Most efficient ways to increase pinning in ceramic superconductors are chemical and thermal modifications of grain morphology, or deliberate introduction of non-superconducting pinning centers. A serious problem and uncertainty arises due to interplay of bulk and grain-boundary pinning and current percolation. Therefore, it is desirable to develop experimental treatments that *simultaneously* enhance intergrain coupling and increase bulk pinning.

This paper describes a novel way to achieve these effects by using heterogeneous sonochemical synthesis. In materials, such as Bi-Sr-Ca-Cu-O, critical current density is mostly determined by the intergrain coupling [2], whereas oxygen content sensitivity of Y-Ba-Cu-O imposes serious constrains on the methods of pinning enhancements. The method is demonstrated on $MgB_2$ superconductor that has two advantages compared to its higher-$T_c$ confreres. First, the irreversible properties of $MgB_2$ are mostly determined by the bulk in-grain pinning forces and, second, it is an s-wave superconductor [3]. The latter is important for the suggested use of ferromagnetic nanoparticles as pinning centers, because s-wave superconductors are much less sensitive to magnetism

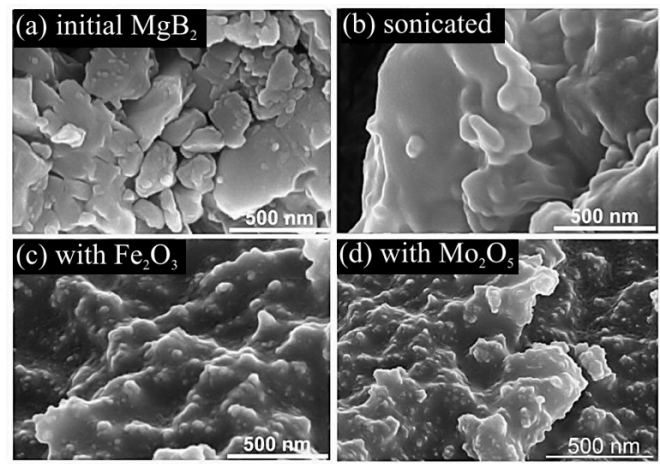

Fig. 1. Scanning electron microscope images of **(a)** initial $MgB_2$ powder; **(b)** sonicated $MgB_2$; **(c)** $MgB_2$ sonicated with $Fe(CO)_5$; **(d)** $MgB_2$ sonicated with $Mo(CO)_6$

compared to *d*-wave cuprates.

Traditional nonmagnetic and nonsuperconducting pinning centers utilize recovery of the condensation energy loss. However, as shown below, if such defects are made ferromagnetic, the net pinning becomes even stronger. The enhancement of pinning by ferromagnetic nanoparticles was demonstrated in conventional superconductors in 60s [4-6] and more recent works studied this effect in thin films and on surfaces of low-$T_c$ superconductors [7-9]. The present work explores distinctly different approach: *bulk* pinning due to finite-size *ferromagnetic nanoparticles* embedded into *ceramic superconductors*. To the best of our knowledge, this was not studied before – neither experimentally, nor theoretically [10-12].

## II. EXPERIMENTAL

### A. Sonochemical Synthesis of Nanocomposites

Acoustic cavitation induced by intense ultrasound produces violent turbulence and shock waves. The implosive collapse of bubbles during cavitation results in extremely high local temperatures (~5000 K), high pressures (~800 atm), but for very short times (< 100 ns) which leads to high cooling rates, about ~ $10^{10}$ K/s [13-15]. Shockwaves stimulate high-velocity collisions between suspended particles with the effective temperatures sufficient to partially melt collided particles, which causes localized inter-particle welding and "neck" formation [13, 14]. The estimated speed of colliding particles approaches half of the speed of sound. In our experiments, $MgB_2$ polycrystalline powder was ultrasonically irradiated for 60 min at -5 °C in 15 ml of decalin (0.1-2 %wt slurry loading, at 20 kHz and ~50 W/cm$^2$) under ambient atmosphere using direct-immersion ultrasonic horn (*Sonics* VCX-750). Similar sets of slurries were sonicated with 1.8 mmol of $Fe(CO)_5$ to produce superconductor – *ferromagnetic nanoparticle* $MgB_2$-$Fe_2O_3$ nanocomposites. Similarly, $MgB_2$ slurries were sonicated with 1.8 mmol of $Mo(CO)_6$ to produce *non-magnetic* $Mo_2O_5$ pinning centers.

Fig. 1 shows scanning electron microscope images of the (a) initial $MgB_2$ powder, (b) sonicated $MgB_2$, (c) $MgB_2$ sonicated with $Fe(CO)_5$ that produced embedded 5-50 nm $Fe_2O_3$ nanoparticles, (d) $MgB_2$ sonicated with $Mo(CO)_6$ resulted in embedded non-magnetic $Mo_2O_5$ nanoparticles. The dramatic change of morphology due to sonication is obvious. Material becomes more compact with fused together grains. Small "knobs" visible in Fig.1 (c) and (d) are $Fe_2O_3$ and $Mo_2O_5$ particles and agglomerates of such particles. To support this statement, local energy dispersive x-ray spectroscopy (EDX) was performed. Fig.2 (a) shows EDX spectra for initial sample and sonicated sample. With the except for small amount of Ti, coming from the abrasion of ultrasonic horn, and increased percentage of oxygen due to sonication under ambient atmosphere, the two spectra are identical. Fig.2 (b) shows spectra collected from nanocomposite with $Fe_2O_3$ nanoparticles. An on-spot spectrum is compared to the off-spot spectrum (locations are shown in the inset). Clearly, the spot (or "knob") contains significant amount of Fe, whereas off-spot area is analogous to that Fig.2(a) showing sonicated $MgB_2$. Similar situation is observed in Fig.2(c) which shows $MgB_2$ with $Mo_2O_5$ nanocomposite. These measurements firmly support our conclusion that described technique a) produces highly compact material with b) nanosized pinning centers embedded into the bulk of ceramic superconductor.

### B. Irreversible Magnetic Properties

To verify that structural modifications result in desirable changes of superconducting properties, magnetic response was measured by using *Quantum Design* MPMS magnetometer. All samples were cold-pressed from powders at 2.5 GPa for 12 hours to form 1.5 mm thick pellets of 6.4 mm in diameter. For measurements $1.5 \times 3 \times 6$ mm$^3$ slabs were cut from the pellets. Fig. 3 shows magnetization loops measured in $MgB_2+Fe_2O_3$ nanocomposite material at 30 K (open symbols) compared to the magnetization loop measured at 42 K. This measurements shows that intrinsic magnetic hysteresis due to magnetic nanoparticles themselves is very small as expected for superparamagnetic material. Also, *M(H)* curve at 42 K is described well by Langevin function. The hysteresis in a superconducting state is significant even at 30

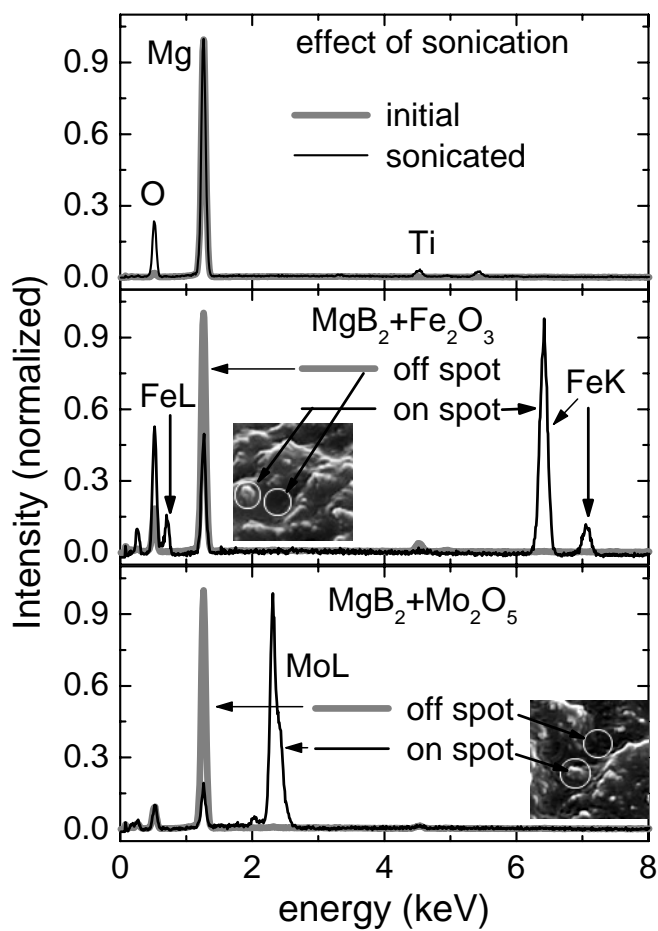

Fig. 2. Local energy dispersive x-ray (EDX) spectra obtained in **(a)** initial and sonicated samples; **(b)** $MgB_2+Fe_2O_3$ nanocomposite – spectra measured on a spot (shown in the inset) compared to the off-spot spectra; **(c)** $MgB_2+Mo_2O_5$ measurements similar to (b) above.

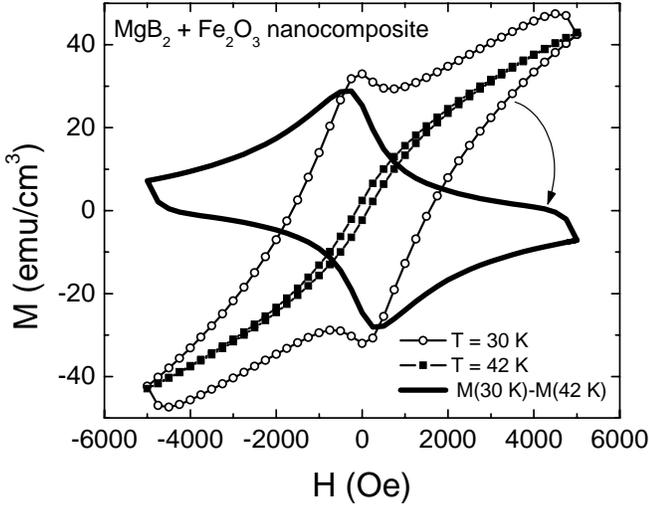

Fig. 3. Magnetization as a function of applied field in MgB$_2$+Fe$_2$O$_3$ nanocomposite measured at 30 K (open circles) shown together with *M(H)* curve measured above $T_c$, at 42 K (solid squares). Solid line is the subtracted magnetization loop.

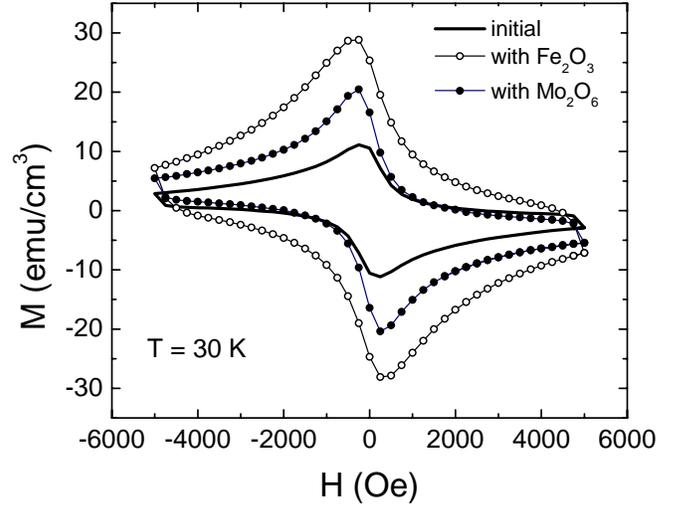

Fig. 4. Comparison of the magnetization loops at 30 K for initial material (solid curve), MgB$_2$+Mo$_2$O$_5$ nanocomposite (solid symbols) and MgB$_2$+Fe$_2$O$_3$ nanocomposite (open symbols).

K. More detailed data including measurements at lower temperatures are shown elsewhere [10, 11]. The solid line in Fig.3 was obtained by a direct subtraction, $M(H,30K) - M(H,42K)$. Indeed, it looks just as regular hysteresis curve of a superconductor.

To compare different samples, the measured curves were normalized by the initial slope, $|dM/dH|_{H\to 0} \approx V/4\pi(1-N)$, measured at 5 K. Here *V* is sample volume and perfect diamagnetism is assumed. Fig. 4 compares magnetization loops measured at 30 K in the initial material (solid curve), MgB$_2$+Mo$_2$O$_5$ nanocomposite (solid symbols) and MgB$_2$+Fe$_2$O$_3$ nanocomposite (open symbols). Evidently, both nanocomposite materials show significant enhancement of the magnetic hysteresis. The hysteresis is almost 300% larger in superconductor containing magnetic Fe$_2$O$_3$ pinning centers, which implies considerable contribution of magnetic pinning. It is important to note that sonication did not affect the superconducting transition temperature, hence chemical properties of the initial material. Details of this work, including dependence of critical current on slurry loading and amount of obtained nanoparticles is given elsewhere [11].

### III. THEORY AND DISCUSSION

#### A. Finite-size ferromagnetic inclusion in a superconductor

Surprisingly, there is no theory describing the interaction of ferromagnetic inclusion of finite size in the superconductor's interior with Abrikosov vortices. Early models treated ferromagnetic nanoparticles as point-like zero-dimensional objects which simply modified the mean-field magnetic induction [4-6]. For the inclusion of finite size, the situation is considerably more difficult due to non-trivial boundary conditions and difficulties of describing the microscopic mechanism of interaction of a ferromagnetic nanoparticle with the vortex core. In this paper we ignore complications of the core interactions and look for join solutions of the London equation in the superconductor bulk and Maxwell equations inside the magnetic particle. Consider a spherical magnetic particle of radius *R* and magnetization $\vec{M}$, embedded into an infinite type-II superconductor containing a single straight vortex line at the distance $\rho$ from the center of the particle. The London equation for the vector-potential $\vec{A}$ ($\nabla \times \vec{A} = \vec{B}$, $\nabla \vec{A} = 0$) in a superconductor, $\vec{A} - \lambda^2 \Delta \vec{A} = 0$ and the Maxwell equation inside the magnetic particle, $\Delta \vec{A} = 0$, can be conveniently solved in spherical coordinates in which vector-potential $\vec{A}$ has only one component $(0, A_\varphi(\rho,\theta), 0)$. The vector potential and tangential component of the magnetic field must be continuous on the particle's surface,

$$\left. \begin{array}{l} A_\varphi^{sc}\big|_{r=R} = A_\varphi^m\big|_{r=R} \\ H_t^{sc}\big|_{r=R} = H_t^m\big|_{r=R} \end{array} \right\} \Rightarrow \left(\nabla \times \vec{A}^{sc}\right)_\theta\big|_{r=R} = \left(\nabla \times \vec{A}^m - 4\pi\vec{M}\right)_\theta\big|_{r=R}$$

Here *m* stands for the solution inside magnetic sphere and *sc* denotes superconductor. In addition, vector potential should vanish inside a superconductor at $\rho \to \infty$ and be finite inside the magnetic sphere. The appropriate solution is:

$$\begin{cases} A_\varphi = \dfrac{4\pi M R \sin\theta}{\left(1+3(\lambda/R)+3(\lambda/R)^2\right)} \dfrac{(1+\rho/\lambda)}{(\rho/\lambda)^2} \exp\left(-\dfrac{(\rho-R)}{\lambda}\right), & \rho \geq R \\ A_\varphi = \dfrac{4\pi M \rho \sin\theta}{\left(1+3(\lambda/R)+3(\lambda/R)^2\right)} \dfrac{(1+R/\lambda)}{(R/\lambda)^2}, & \rho < R \end{cases}$$

The corresponding supercurrent density induced by the particle at the distance $\rho$, $4\pi c^{-1} j_\varphi = -A_\varphi \lambda^{-2}$ is:

$$j_\varphi = -\dfrac{cMR}{\left(1+3(\lambda/R)+3(\lambda/R)^2\right)} \dfrac{(1+\rho/\lambda)}{\rho^2} \exp\left(-\dfrac{\rho-R}{\lambda}\right) \sin\theta$$

#### B. Magnetic pinning force

Knowing the current density distribution, the magnetic pinning force can be calculated from:

$$\vec{f}_{mag} = c^{-1} \int \left[ \vec{j}(\rho_v, \theta_v) \times \vec{\Phi}_0 \right] dl$$

where integration is curried along the entire vortex. In a general case of an arbitrary orientation, the value of magnetic pinning force is scaled by the factor of $\cos(\alpha)$, where $\alpha$ is the angle of misalignment. In addition, the vortex is experiencing additional moment of forces which is trying to align it along the magnetic moment of the sphere [10]. Results of calculated total *magnetic* pinning force is shown in Fig.5 for different sizes of ferromagnetic sphere compared to the penetration depth. The Ginsburg-Landau parameter was chosen $\kappa = 100$, close to the experimental values for MgB$_2$[3]. As follows from the calculations, the force experienced by an Abrikosov vortex is attractive and, at larger distances exponentially depends on the separation from the magnetic particle. Compared to the short-range core pinning (affective at distance of the order of the coherence length), magnetic pinning force is long-range (characteristic length scale is the London penetration depth). Since all ferromagnetic nanoparticles also contribute the usual core pinning, such nanosized ferromagnetic nanoparticles are ideal pinning centers.

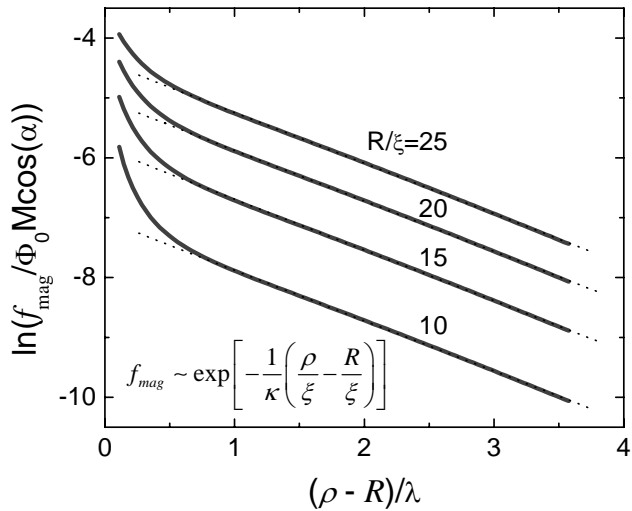

Fig. 5. Magnetic pinning force as a function of the distance between a vortex and a magnetic sphere calculated for spheres of different size.

### C. Extension to high-$T_c$ superconductors

The main technological difficulty of preparing described nano-composite superconducting materials is embedding of ferromagnetic nanoparticles into the bulk of ceramic superconductors. In the case of high-$T_c$ cuprates such as BSCCO and YBCO this issue is crucial, because these materials form variety of phases depending on crystallization conditions. Therefore, direct mixing of nanoparticles into the melt is not feasible (not to mention very large surface tension between the melt and, for example, Fe$_2$O$_3$ nanoparticles). Heterogeneous sonochemical synthesis avoids these difficulties by preparing ferromagnetic nanoparticles *in situ* during the reaction. A more difficult problem is the d-wave nature of high-$T_c$ cuprates. These unconventional superconductors are much more sensitive to magnetism and, therefore, possible contamination of superconducting material with iron during synthesis must be taken into account and the process should be optimized to avoid this effect. One of the ways is to use various gases during synthesis. For example, excess of oxygen binds iron before it has chance to enter the crystal lattice of a superconductor.

Nevertheless, direct effect of sonication on the morphology of high-$T_c$ granular materials is dramatic. Fused together grains and more homogeneous material, indeed, lead to improved superconducting properties. Our recent results show significant enhancement of magnetic and *transport* properties in sonicated BSCCO-2212 [12]. The method has also been adopted for oxygen – sensitive YBCO and yielded promising results.

### ACKNOWLEDGMENT

Discussions with K. S. Suslick, V. Geshkenbein, A. A. Polyanskii, A. Gurevich, and B. Ivlev are greatly appreciated.

### REFERENCES

[1] A. M. Campbell and J. E. Evetts, *Critical currents in superconductors*. London: Taylor & Francis Ltd., 1972.
[2] D. Larbalestier, A. Gurevich, D. M. Feldmann, and A. Polyanskii, "Superconductors: Pumping up for wire applications," *Nature*, vol. 414, pp. 368-377, 2001.
[3] Review, "Superconductivity in MgB$_2$: electrons, phonons and vortices," *Physica C*, vol. 385, pp. 1-305, 2003.
[4] T. H. Alden and J. D. Livingston, "Magnetic pinning in a type-II superconductor," *Appl. Phys. Lett.*, vol. 8, pp. 6-7, 1966.
[5] T. H. Alden and J. D. Livingston, "Ferromagnetic particles in a type-II superconductor," *J. Appl. Phys.*, vol. 37, pp. 3551-6, 1966.
[6] C. C. Koch and G. R. Love, "Superconductivity in niobium containing ferromagnetic gadolinium or paramagnetic yttrium dispersions," *J. Appl. Phys.*, vol. 40, pp. 3582-7, 1969.
[7] Y. Nozaki, Y. Otani, K. Runge, H. Miyajima, and B. Pannetier, "Magnetostatic effect on magnetic flux penetration in superconducting Nb film covered with a micron-size magnetic particle array," *J. Appl. Phys.*, vol. 79, pp. 6599-6601, 1996.
[8] M. J. V. Bael, K. Temst, V. V. Moshchalkov, and Y. Bruynseraede, "Magnetic properties of submicron Co islands and their use as artificial pinning centers," *Phys. Rev. B*, vol. 59, pp. 14674 -14679, 1999.
[9] J. I. Martin, M. Velez, J. Nogues, and I. K. Schuller, "Flux Pinning in a Superconductor by an Array of Submicrometer Magnetic Dots," *Phys. Rev. Lett.*, vol. 79, pp. 1929–1932, 1997.
[10] A. Snezhko, T. Prozorov, and R. Prozorov, "Magnetic nanoparticles as efficient bulk pinning centers in type-II superconductors," *cond-mat/0403015, submitted for publication*, 2004.
[11] T. Prozorov, R. Prozorov, A. Snezhko, and K. S. Suslick, "Sonochemical Modification of the Superconducting Properties of MgB$_2$," *Appl. Phys. Lett.*, vol. 83, pp. 2019-2021, 2003.
[12] T. Prozorov, B. McCarty, Z. Cai, R. Prozorov, and K. S. Suslick, "Effects of High Intensity Ultrasound on BSCCO-2212 Superconductor," *Appl. Phys. Lett.*, vol. in print, 2004.
[13] K. S. Suslick and G. J. Price, "Applications of Ultrasound to Materials Chemistry," *J. Ann. Rev. Mat. Sci.*, vol. 29, pp. 295-326, 1999.
[14] K. S. Suslick and S. J. Doctycz, "Interparticle collisions driven by ultrasound," *Science*, vol. 247, pp. 1067-1069, 1990.
[15] M. Gasgnier, L. Albert, J. Derouet, and L. Beaury, "Ultrasound effects on various oxides and ceramics: macro- and microscopic analyses," *J. Solid State. Chem.*, vol. 115, pp. 532-9, 1995.